\documentclass[iop]{emulateapj}
\usepackage{apjfonts}

\usepackage{epsfig,graphics,graphicx}
\pdfoutput=1

\newcommand{\arcm}{\hbox{$^\prime$}}

\newcommand{\degree}{\hbox{$^\circ$}}
\newcommand{\rosat}{\emph{ROSAT}}
\newcommand{\chandra}{\emph{Chandra}}
\newcommand{\xmm}{\emph{XMM-Newton}}
\newcommand{\xmms}{\emph{XMM}}

\newcommand{\arcs}{\mbox{\arcm\arcm}}
\newcommand{\Zsol}{\ensuremath{\mathrm{~Z_{\odot}}}}

\newcommand{\Msol}{\ensuremath{\mathrm{~M_{\odot}}}}
\newcommand{\Msolpyr}{\ensuremath{\mathrm{~M_{\odot}~yr^{-1}}}}

\newcommand{\NH}{\ensuremath{N_{\mathrm{H}}}} 

\newcommand{\s}{\ensuremath{\mathrm{~s}}}
\newcommand{\ps}{\ensuremath{\s^{-1}}}
\newcommand{\cm}{\ensuremath{\mathrm{~cm}}}
\newcommand{\pcmsq}{\ensuremath{\cm^{-2}}}
\newcommand{\pcmcu}{\ensuremath{\cm^{-3}}}
\newcommand{\km}{\ensuremath{\mathrm{~km}}}
\newcommand{\Mpc}{\ensuremath{\mathrm{~Mpc}}}
\newcommand{\pMpc}{\ensuremath{\Mpc^{-1}}}
\newcommand{\kmpspMpc}{\ensuremath{\km \ps \pMpc\,}}
\newcommand{\erg}{\ensuremath{\mathrm{~erg}}}
\newcommand{\ergps}{\ensuremath{\erg \ps}}

\newcommand{\ergpcmcu}{\ensuremath{\erg \pcmcu}}
\newcommand{\kmps}{\ensuremath{\km \ps}}

\newcommand{\Hi}{\ensuremath{\mathrm{H}\textsc{i}}}

\newcommand{\Dtf}{\ensuremath{D_{25}}}

\begin{document}

\submitted{}
\received{2014 June 16}
\accepted{2014 July 26}

\title{Deep Chandra Observations of HCG~16 --- II. The Development of the Intra-group Medium in a Spiral-Rich Group}
\author{E. O'Sullivan\altaffilmark{1}, J.~M. Vrtilek\altaffilmark{1}, L.~P. David\altaffilmark{1}, S. Giacintucci\altaffilmark{2,3}, A. Zezas\altaffilmark{1,4}, T.~J. Ponman\altaffilmark{5}, G.~A. Mamon\altaffilmark{6},\\ P. Nulsen\altaffilmark{1} and S. Raychaudhury\altaffilmark{7,5}}
\altaffiltext{1}{Harvard-Smithsonian Center for Astrophysics, 60 Garden
  Street, Cambridge, MA 02138, USA}
\altaffiltext{2}{Department of Astronomy, University of Maryland, College
  Park, MD 20742-2421, USA}
\altaffiltext{3}{Joint Space Science Institute, University of Maryland, College Park, MD 20742-2421, USA}
\altaffiltext{4}{Physics Department and Institute of Theoretical \& Computational Physics, University of Crete, GR-71003 Heraklion, Crete, Greece}
\altaffiltext{5}{School of Physics and Astronomy, University of Birmingham, Birmingham, B15 2TT, UK}
\altaffiltext{6}{Institut d'Astrophysique de Paris (UMR 7095 CNRS \& UMPC), 98 bis Bd Arago, F-75014 Paris, France}
\altaffiltext{7}{Department of Physics, Presidency University, 86/1 College Street, 700073 Kolkata, India}
\shorttitle{The development of the intra-group medium in HCG~16}
\shortauthors{O'Sullivan et al}

\begin{abstract}
  We use a combination of deep \chandra\ X-ray observations and radio
  continuum imaging to investigate the origin and current state of the
  intra-group medium in the spiral-rich compact group HCG~16. We confirm
  the presence of a faint ($L_{X,{\rm
      bolo}}$=1.87$^{+1.03}_{-0.66}$$\times$10$^{41}$\ergps), low
  temperature (0.30$^{+0.07}_{-0.05}$~keV) intra-group medium (IGM)
  extending throughout the ACIS-S3 field of view, with a ridge linking the
  four original group members and extending to the southeast, as suggested
  by previous \rosat\ and \xmm\ observations. This ridge contains
  6.6$^{+3.9}_{-3.3}$$\times$10$^9$\Msol\ of hot gas and is at least partly
  coincident with a large-scale \Hi\ tidal filament, indicating that the
  IGM in the inner part of the group is highly multi-phase. We present
  evidence that the group is not yet virialised, and show that gas has
  probably been transported from the starburst winds of NGC~838 and NGC~839
  into the surrounding IGM. Considering the possible origin of the IGM, we
  argue that material ejected by galactic winds may have played a
  significant role, contributing 20-40\% of the observed hot gas in the
  system.

\end{abstract}


\section{Introduction}

The majority of galaxies in the local universe reside in small,
gravitationally bound groups \citep{Ekeetal04}, whose low velocity
dispersions ($\lesssim$500\kmps) and small galaxy separations are conducive
to tidal interactions and mergers between group members. X-ray observations
have shown that many groups host extended halos of hot gas, but that the
existence of a hot intra-group medium (IGM) appears to be linked to the
presence of early-type galaxies. X-ray luminous groups are sometimes
described as miniature galaxy clusters; their IGM is highly enriched,
particularly in the group core, they host significant early-type galaxy
populations, are often dominated by a single, centrally-located giant
elliptical, and follow morphology-density and morphology-radius relations
similar to (but offset from) those seen in clusters
\citep{HelsdonPonman03a}. However, they are typically poor in cold gas,
with neutral hydrogen restricted to spiral galaxies in the group outskirts
\citep{Kilbornetal06}.

Conversely, spiral-rich groups are typically poor in hot gas, and
consequently X-ray faint. \citet{Mulchaeyetal03} find that spiral-rich
groups tend to be less X-ray luminous than their elliptical-dominated
cohorts, and detect none of the twelve spiral-only groups in their \rosat\
atlas of 109 systems. \citet{OsmondPonman04} detect only galaxy-scale
emission in the ten spiral-only groups in the GEMS sample; the only
possible exception, the NGC~3783 group, appears to be biased by the
exceptionally luminous Seyfert nucleus of the dominant spiral. However,
spiral-rich groups usually contain significant quantities of cold gas in and
around their galaxies. Examination of X-ray faint, spiral-rich compact
groups has led to the suggestion of an evolutionary sequence, with galaxy
interactions stripping the \Hi\ from spiral galaxies to form intergalactic
clouds and filaments or even a diffuse cold IGM
\citep{VerdesMontenegroetal01,Johnsonetal07,Konstantopoulosetal10}. The
redistribution of the \Hi\ component is accompanied by the transformation
of some member galaxies from late to early-type, and in some cases by star
formation.

While the role of tidal interactions in driving the evolution of the galaxy
population and \Hi\ component is clear, the origin of the hot IGM and its
link to the development of the group is not. Infall and gravitational shock
heating is believed to be the primary source of the hot gas which makes up
the dominant baryonic component of massive clusters, and the same mechanism
probably provides most of the IGM in the most massive groups. However, the
connection between galaxy evolution and the presence of a hot IGM in low
mass groups suggests a link. One possibility is that star-formation-driven
galactic winds could contribute to the formation of the IGM. Another is
that intergalactic \Hi\ could be shock heated by collisions within the
group. The latter process is observed in one system, Stephan's Quintet
(HCG~92), in which an infalling spiral galaxy has collided with a tidal \Hi\
filament, heating it to a temperature of $\sim$0.6~keV
\citep{VanderhulstRots81,Sulenticetal01,Trinchierietal03,OSullivanetal09}.

Understanding the development of the hot IGM is clearly central to any
study of galaxy evolution in groups, or of structure formation involving
groups. However, relatively few suitable systems have been examined in
detail, and their X-ray faintness makes such examination challenging. Only
two groups at an earlier evolutionary stage than Stephan's Quintet have
been shown to possess a hot IGM using modern high-spatial-resolution X-ray
instruments. \citet{Trinchierietal08} confirmed the presence of diffuse hot
gas in SCG0018-4854, a spiral-only southern compact group of four galaxies,
but the short \xmms\ observation provided only limited information on the
state and origin of the IGM and its relation to the galaxy population. The
other example is the well known spiral-dominated group HCG~16, which we
have chosen to study in this paper. HCG~16 appears to contain significant
quantities of diffuse hot and cold gas (\Hi), and its galaxy members
include starburst systems with outflowing winds which may be enriching
their surroundings. The group therefore appears to be at the start of the
process of galaxy merger, hot gas build-up and enrichment which will
eventually produce an X-ray bright, metal-rich IGM and an
elliptical-dominated group.

In order to examine the process of IGM development in HCG~16, we have
obtained deep \chandra\ X-ray observations of the group, as well as new
\textit{Giant Metrewave Radio Telescope} (GMRT) 610~MHz radio data. In this
paper we combining these observations with archival \chandra\ and
\textit{Very Large Array} (VLA) 1.4~GHz data, with the goal of determining
the physical properties and origin of the diffuse gas component, and its
relationship to the galaxy population. We describe the observations, data
reduction and analysis techniques in detail in \citet[hereafter Paper
I]{OSullivanetal14b_special}. All five of the major galaxies in the group show
evidence of star formation and/or nuclear activity, with two of the
galaxies hosting galactic superwinds. A full discussion of the properties
of the galaxies and their point source populations can be found in Paper I,
and we summarise the results relevant to the current paper in
Section~\ref{sec:winds}.

Throughout this paper we adopt a redshift of $z$=0.0132 for the group
\citep{Hicksonetal92} and a Galactic hydrogen column density of
\NH=2.56$\times$10$^{20}$\pcmsq\ for the four original group member
galaxies and the surrounding diffuse emission \citep[taken from the
Leiden/Argentine/Bonn survey,][]{Kalberlaetal05}. For NGC~848 we adopt a
hydrogen column of \NH=2.75$\times$10$^{20}$\pcmsq. All fluxes and
luminosities are corrected for Galactic absorption. A redshift-independent
distance measurement is available for one of the five major galaxies, a
Tully-Fisher distance of 56.5~Mpc for NGC~848 \citep{Theureauetal07}. This
is consistent, within errors, with redshift-based estimates for all five of
the galaxies, correcting for infall toward the Virgo cluster, great
attractor and Shapley Supercluster, for a cosmology with $H_0$=70\kmpspMpc.
We therefore adopt this distance estimate for the group as a whole, which
gives an angular scale of 1\arcs=273~pc.

\section{HCG~16}
\label{sec:review}

\begin{figure*}
\centerline{
\includegraphics[width=1.05\columnwidth,bb=36 175 576 616]{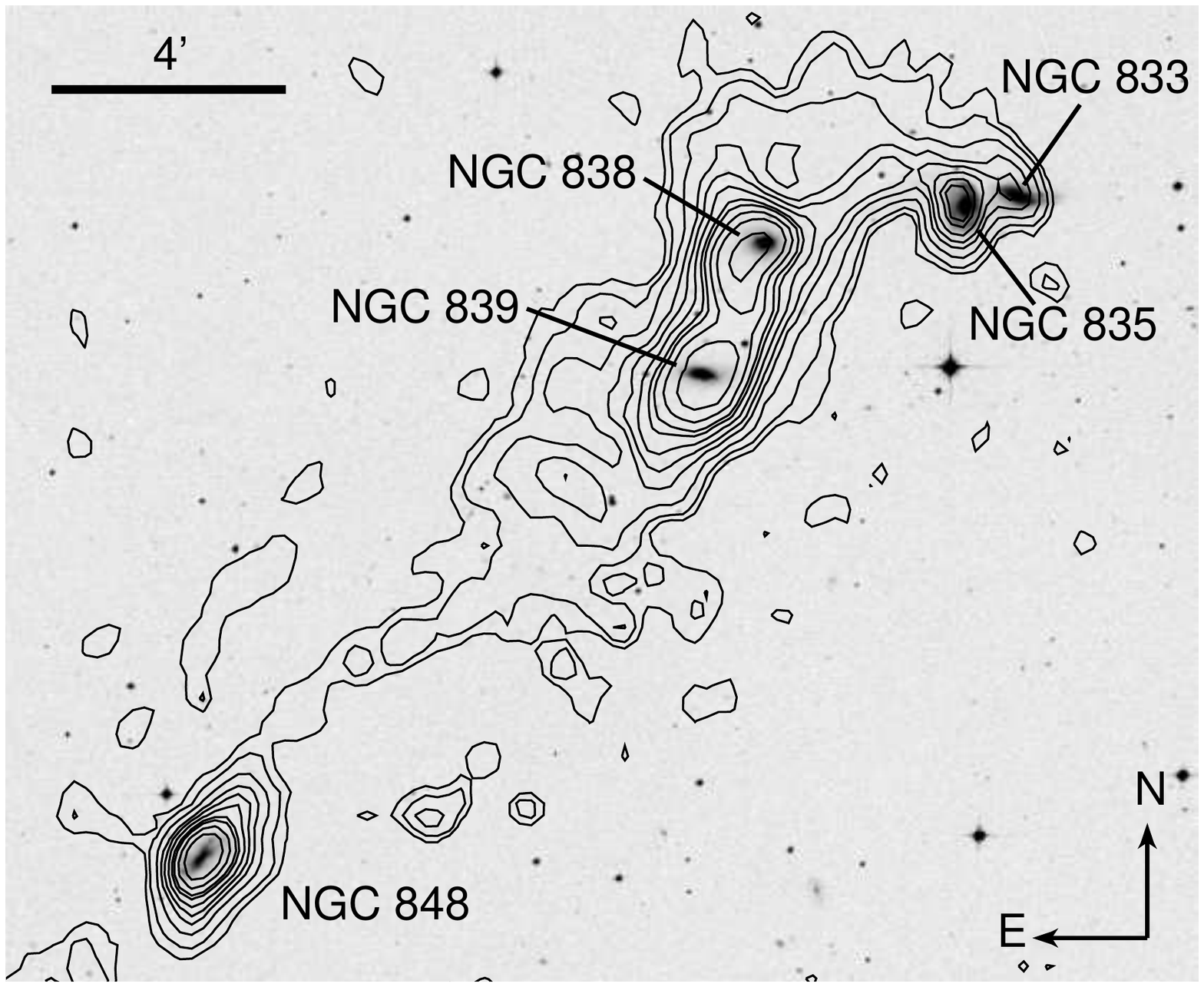}
\includegraphics[width=1.05\columnwidth,bb=36 175 576 616]{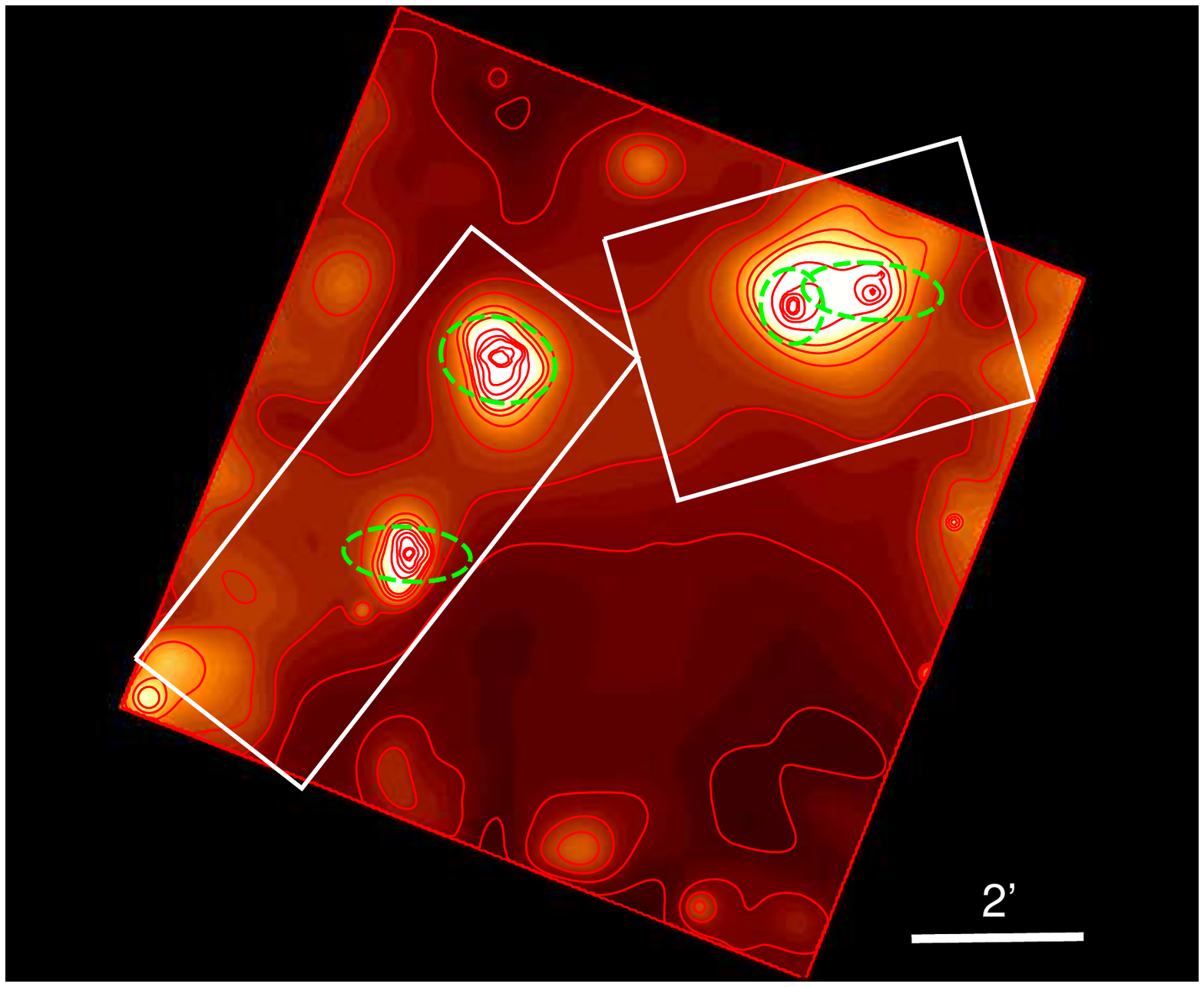}
}
\centerline{
\includegraphics[width=1.05\columnwidth,bb=36 175 576 616]{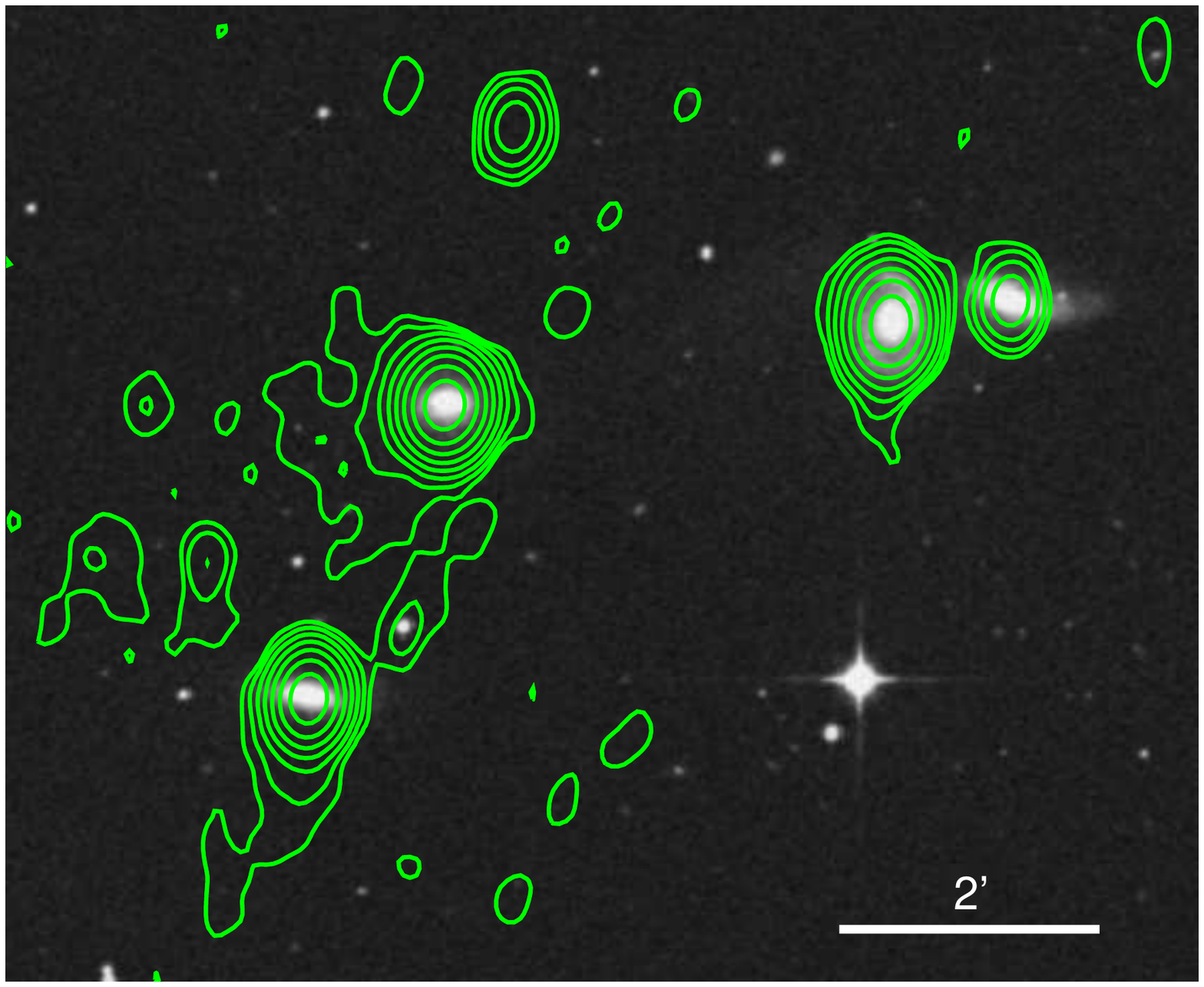}
\includegraphics[width=1.05\columnwidth,bb=36 175 576 616]{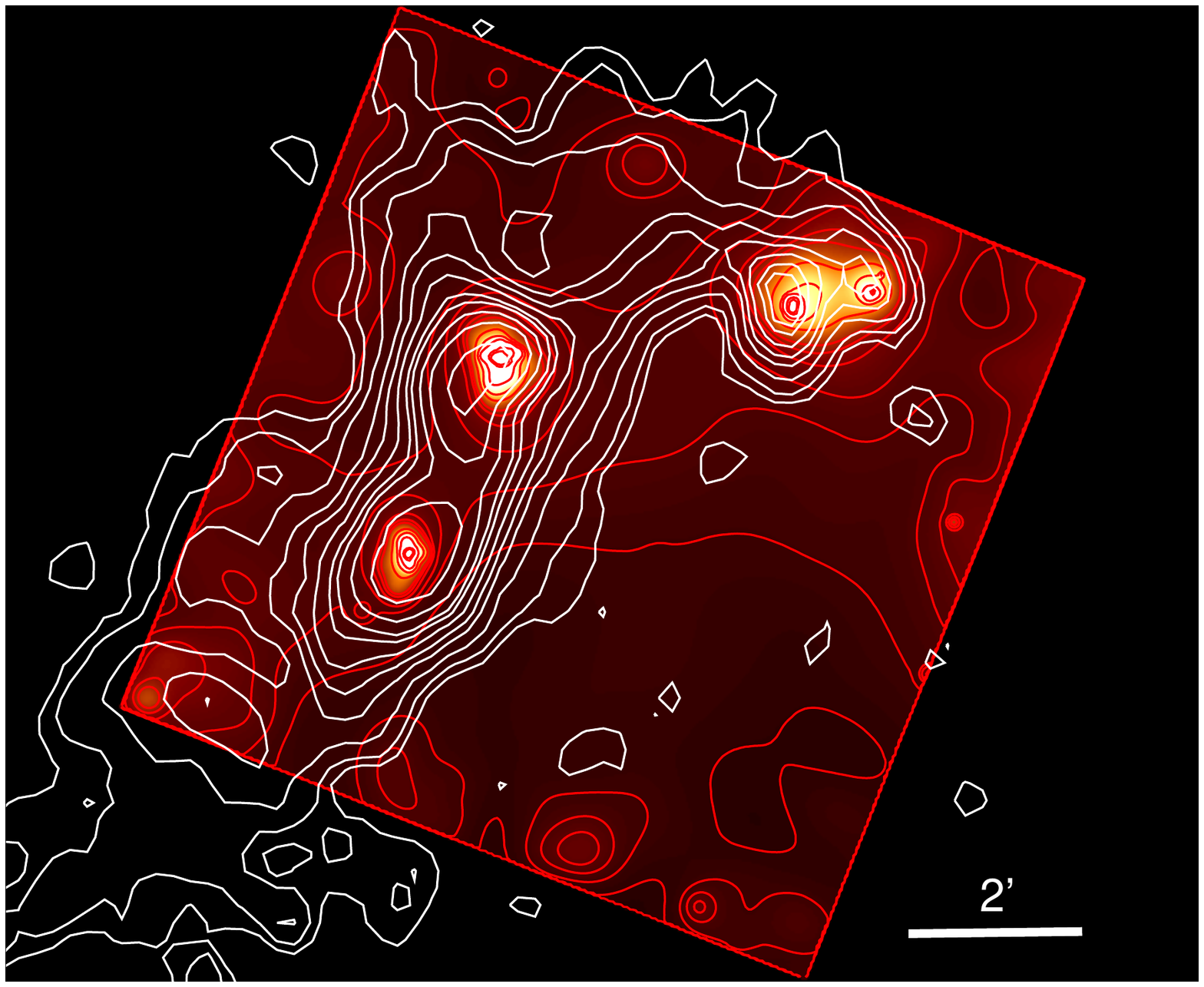}
}
\caption{\label{fig:Xopt} \textit{Upper left:} Digitized Sky Survey 2 (DSS2) $R$-band image of the five largest galaxies in HCG~16, with the four galaxies originally identified as a compact group to the northwest. VLA \Hi\ contours from Verdes-Montenegro et al. (2014, in prep.) are overlaid, with levels N(\Hi)$\simeq$10,20,40,65,85,110,140,160,200,250,350,450,570$\times$10$^{-19}$\pcmsq. \textit{Upper right:} Adaptively smoothed \chandra\ 0.5-2~keV image using data from the S3 CCD in all five observations. Contours are overlaid in red to help elucidate the distribution of diffuse emission. Dashed ellipses indicate the \Dtf\ contours of the four main galaxies, white boxes the spectral extraction regions used to characterize the ridge of diffuse emission. \textit{Lower left:} VLA 1.4~GHz contours overlaid on a DSS2 image of the four main galaxies. Contours start at 3$\times$ the 0.2~mJy~beam$^{-1}$ rms noise level and increase in steps of factor 2. \textit{Lower right:} VLA \Hi\ contours overlaid on the adaptively smoothed \chandra\ 0.5-2~keV image. All four images have the same orientation.}
\end{figure*}

HCG~16, also known as Arp~318, was originally identified \citep{Hickson82}
as a compact group of four spiral galaxies, NGC~833 (HCG~16B), NGC~835 (A),
NGC~838 (C) and NGC~839 (D), with later studies identifying a fifth large
spiral galaxy member (NGC~848) and a surrounding halo of dwarf galaxies
\citep[e.g.,][]{Ribeiroetal98}.  Figure~\ref{fig:Xopt} includes a Digitized
Sky Survey image showing the relative positions of the group members. All
five major galaxies host AGN and/or starbursts
\citep[e.g.,][]{Martinezetal10,DeCarvalhoCoziol99,Continietal98}, and tidal
structures suggest an an ongoing or recent interaction between NGC~833 and
NGC~835 \citep{Konstantopoulosetal13}.

Neutral hydrogen mapping of the group revealed a $\sim$20\arcm\ long
complex filament of cold gas surrounding the four original members of the
group and linking them to NGC~848 \citep{VerdesMontenegroetal01}, almost
certainly as the result of tidal interactions between group members. The
total mass of \Hi\ in the group is $>$2.63$\times$10$^{10}$\Msol, and
Verdes-Montenegro et al. estimate that the group is $<$30\% \Hi-deficient.
The four original member galaxies are $\sim$50-80\% deficient, while
NGC~848 is only $\sim$7\% deficient. This suggests that the majority of the
intergalactic \Hi\ originated in the four main galaxies, perhaps being
transported out into the IGM by interactions among them, and then drawn
into its current morphology by a close passage of NGC~848.
\citet{Borthakuretal10} show that the \Hi\ velocity distribution covers the
range $\sim$3650-4100\kmps, confirming its association with the major
member galaxies.

\rosat\ studies of HCG~16 in the X-ray band were able to separate emission from the galaxies and diffuse inter-galactic gas, finding a gas temperature $\sim$0.3~keV and tracing emission out to $\sim$8\arcm\ \citep{Ponmanetal96,DosSantosMamon99}. \rosat\ imaging found the gas distribution to be irregular, with the brightest emission around the four original member galaxies and to the southeast of NGC~839, with fainter extension west or southwest of NGC~833 \citep{DosSantosMamon99}. 

First light data from \xmm\ were used to investigate further the gas distribution, but this was hampered by a combination of uncertain calibration and scattered light from a bright background source close to the edge of the field of view. Nonetheless \citet{Belsoleetal03} reported a highly elliptical diffuse emission component surrounding the four main galaxies, with a temperature of $\sim$0.5~keV and abundance $\sim$0.07\Zsol. A short (12.5~ks) \chandra\ observation in cycle~1 provided no additional evidence of group-scale emission \citep{Jeltemaetal08}.

Optical spectroscopic studies of the starburst galaxies NGC~838 and NGC~839
have provided a detailed characterization of their stellar structures and
outflowing galactic winds. In NGC~839 the wind has formed a biconical polar
outflow \citep{Richetal10}, while in NGC~838 wind outflows have inflated
bubbles north and south of the galaxy, above and below the galactic disk
\citep{Vogtetal13}. These bubbles are likely confined by the surrounding
IGM and \Hi, although there is some indication that material is leaking
from the southern bubble. Examination of optical spectra of the stellar
populations suggests that star formation peaked in NGC~838 and NGC~839 500
and 400~Myr ago respectively \citep{Vogtetal13,Richetal10} and still
continues in the core of each galaxy. Our own stellar population modelling
confirms that star formation is ongoing in NGC~838, but shows that star
formation in AGN-dominated NGC~833 has been minimal ($<$3\Msolpyr) over the
past few hundred Myr.

\section{Observations and Data Reduction}
\label{sec:obs}

Paper I describes our observations and reduction techniques in detail. We
used the five available \chandra\ observations of HCG~16, all of which were
made with the ACIS-S3 CCD at the focal point. Three of the observations
(ObsIDs~\dataset[ADS/Sa.CXO#obs/15181]{15181}, \dataset[ADS/Sa.CXO#obs/15666]{15666} and \dataset[ADS/Sa.CXO#obs/15667]{15667}) totaling $\sim$75.7~ks, were made in 2013
July, using the full CCD and the same roll angle. The earliest
observation, ObsID~\dataset[ADS/Sa.CXO#obs/923]{923}, was made in 2000 Nov, for $\sim$12.5~ks. In 2008
Nov (ObsID~\dataset[ADS/Sa.CXO#obs/10394]{10394}) the group was observed in 1/2 subarray for $\sim$13.8~ks.
ObsID 923 and the 2013 observations all cover the four original member
galaxies, but ObsID 10394 only covers NGC~835, NGC~838 and part of NGC~833.

All five pointings were reduced using CIAO 4.6.1
\citep{Fruscioneetal06} and CALDB 4.5.9 following techniques similar to
those described in \citet{OSullivanetal07} and the \chandra\ analysis
threads\footnote{http://asc.harvard.edu/ciao/threads/index.html}. Point
sources were identified using the \textsc{ciao} task \textsc{wavdetect},
and excluded. Spectra were extracted from each dataset using the
\textsc{specextract} task. Abundances were measured relative to the
abundance ratios of \citet{GrevesseSauval98}. 1$\sigma$ errors are reported
for all fitted values.

Reduction and analysis of the GMRT and VLA data was performed in the NRAO
Astronomical Image Processing System (\textsc{aips}) package following the
standard procedure (Fourier transform, clean and restore). Phase-only
self-calibration was applied to remove residual phase variations and
improve the quality of the images. The final VLA (GMRT) image has an
angular resolution of 25$\times$18.1\arcs\ (5.6$\times$5.4\arcs) and an rms
noise level (1$\sigma$) of 0.2 (0.06)~mJy beam$^{-1}$.

\section{Results}
We initially examined X-ray images of the group to determine the basic
structures associated with the galaxies and whether any large scale
emission was visible. Heavily smoothed or binned images showed evidence of
emission between the four galaxies on the ACIS-S3 CCD, located in a ridge
connecting the galaxies. Brighter diffuse emission was also visible in the
disks of several of the galaxies, between NGC~833 and NGC~835, and in
regions north and south of the disks of NGC~838 and NGC~839.

Figure~\ref{fig:Xopt} shows the core of the group, imaged in the optical,
\Hi, 1.4~GHz radio continuum, and soft (0.5-2~keV) X-ray bands. The X-ray
image has been processed to remove point sources outside the galaxy cores
and refill the resulting holes using the \textsc{dmfilth} tool. The image
has then been adaptively smoothed using the \textsc{csmooth} task with
signal-to-noise limits of 3-5$\sigma$, and exposure corrected using a
0.91~keV exposure map smoothed to the same scales. The energy of the
exposure map was chosen to match the modal event energy in the
observations.

The ridge of faint diffuse emission linking the galaxies is clear, and
appears to extend past NGC~839 to the southeast, while surface brightness
declines to the southwest and northeast. The apparent brightness of
features close to the edges of the field may be affected by the exposure
correction and adaptive smoothing processes, and the surface brightness of
the diffuse emission is low. However, as mentioned above, heavily binned
images show the same basic ridge structure.

Comparison with the \Hi\ map shows that the hot and cold gas structures are
similar, with the \Hi\ filament overlapping the X-ray ridge over most of
its length. This is most obvious around NGC~838 and NGC839, the brightest
diffuse X-ray sources in the group, which appear to be embedded in some of
the highest density \Hi\ gas. The extension of the X-ray ridge southeast of
NGC~839 also overlaps the section of the \Hi\ filament extending toward
NGC~848. The \Hi\ and X-ray emission agree less well between NGC838 and
NGC~835, with the X-ray ridge directly linking the two galaxies, while the
\Hi\ filament curls to the north through a region of lower X-ray surface
brightness.

\subsection{Radio and X-ray imaging of the galactic superwinds}
\label{sec:winds} 
We examined the galactic superwinds of NGC~838 and NGC~839 in detail in
Paper I, but as their outflows are possible contributors of hot gas to the
intra-group medium, we summarise the relevant results below, and then discuss
the wind morphology in more detail.

Our spectroscopic analysis of the X-ray and radio properties of the
galaxies confirms that NGC~833 and NGC~835 are dominated by emission from
absorbed AGN, while NGC~838 and NGC~839 are primarily starburst systems,
with only minimal AGN contribution to their X-ray luminosity. We estimate
the star-formation rate in NGC~838 (NGC~839) to be $\sim$7-17\Msolpyr\
($\sim$8-20\Msolpyr), with the lower, infra-red derived values probably
more representative of the current rate. The galactic winds have
temperatures $\sim$0.8~keV, with some evidence of a temperature decline in
the outer part of the southern bubble of NGC~838. We estimate the rate of
outflow in the winds to be 2.5\Msolpyr\ in NGC~839 and $\sim$17\Msolpyr\ in
NGC~838. However, since the NGC~838 wind appears to be largely confined, it is
unclear how much of this material escapes into the surrounding IGM.

\begin{figure}
\includegraphics[width=\columnwidth,bb=36 175 576 616]{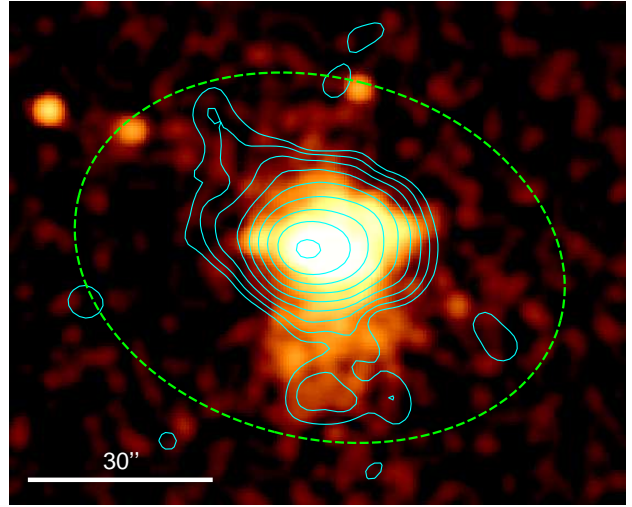}
\caption{\label{fig:N838}\chandra\ 0.5-2~keV exposure corrected image of
  NGC~838, overlaid with GMRT 610~MHz contours. North is at the top of the
  image, west to the right. The image has been smoothed with a
  $\sim$2.5\arcs\ Gaussian. Contours begin at 180~$\mu$Jy~beam$^{-1}$
  (3$\times$ the rms noise level) and increase in steps of factor 2. The
  dashed green ellipse shows the approximate \Dtf\ contour of the galaxy
  stellar light, and the scalebar indicates 30\arcs\ ($\sim$8~kpc).}
\end{figure}

\begin{figure}
\includegraphics[width=\columnwidth,bb=36 175 576 616]{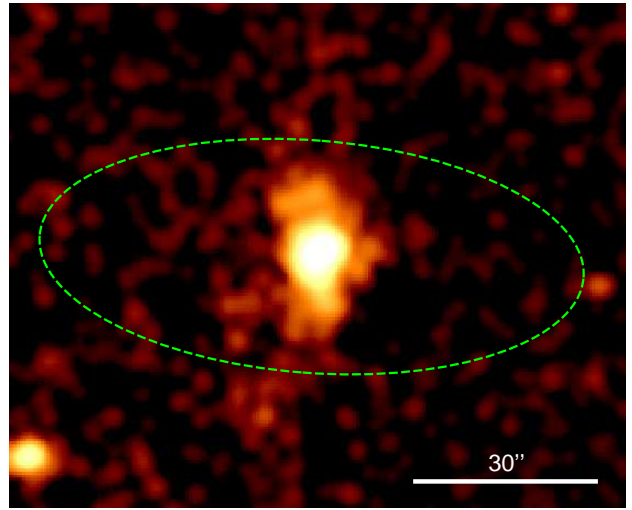}
\caption{\label{fig:N839}\chandra\ 0.5-2~keV exposure corrected image of
  NGC~839. The image has been smoothed with a $\sim$2.5\arcs\ Gaussian. The
  dashed green ellipse shows the approximate \Dtf\ contour of the galaxy
  stellar light, and the scalebar indicates 30\arcs\ ($\sim$8~kpc).}
\end{figure}

Figures~\ref{fig:N838} and \ref{fig:N839} show X-ray images of the two galaxies. In both cases, a central area of X-ray emission, corresponding to the galaxy core, is visible. In the case of NGC~838 this is extended to the east and west in an ellipse. This includes emission from the northern wind bubble, and as the northern side of the galaxy disk is tilted toward the viewer, the galaxy centre is partly screened by this bubble and by gas and dust in the disk. More diffuse emission is clearly visible to the south, extending $\sim$25\arcs\ with a morphology suggestive of a bubble. Comparison with the H$\alpha$ imaging of the galaxy wind by \citet{Vogtetal13} confirms that the X-ray and optical morphology is similar, and that the southern bubble is viewed through the disk of the galaxy. Although the 610~MHz emission is dominated by the galaxy core, it is clearly extended to the south, coincident with the X-ray/optical bubble. Fainter diffuse X-ray and radio emission is visible north of the galaxy core, although interestingly both bands suggest that material outside the northern bubble is extended toward the northeast, rather than continuing the north-south axis of the bubbles.

\begin{figure}
\includegraphics[width=\columnwidth,bb=30 210 570 740]{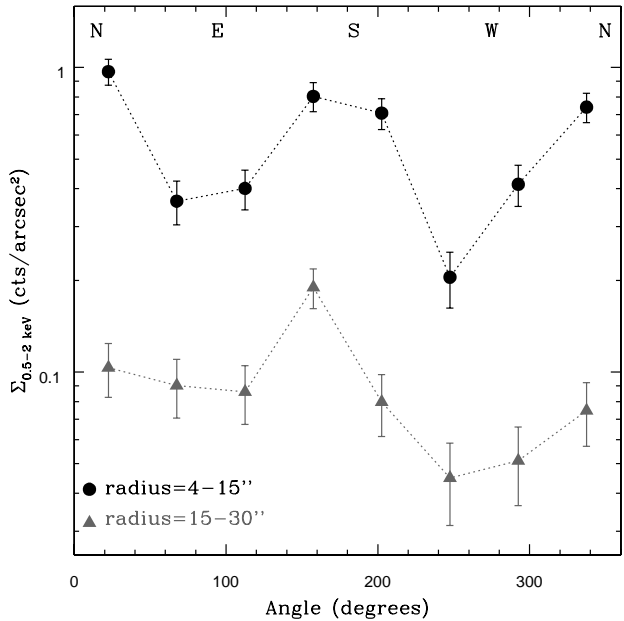}
\caption{\label{fig:wedges}0.5-2~keV surface brightness around NGC~839 in the combined 2013 observations, measured in 45\degree\ sectors of annuli with radii 4-15\arcs\ (black circles) and 15-30\arcs\ (grey triangles). Error bars indicate 1$\sigma$ uncertainties.}
\end{figure}

NGC~839 is a point source at 610~MHz, dominated by the emission from its
dense star-forming core. This is also bright in the X-ray band, but it is
clear that gas emission extends north and south of the core, with a
morphology comparable to the biconical outflow observed in H$\alpha$
\citep{Richetal10}. However, there is also an indication of fainter X-ray
emission extending to the southeast. To test whether this apparent
extension is real, we measure the number of 0.5-2~keV counts in two annuli,
one with radius 4-15\arcs\ ($\sim$1-4~kpc) corresponding to the brighter conical wind
regions, one with radius 15-30\arcs\ ($\sim$4-8~kpc) corresponding to the
possible fainter emission. We break each annulus into 45\degree\ sectors,
starting from north and proceeding anti-clockwise. Figure~\ref{fig:wedges}
shows the resulting azimuthal surface brightness measurements. It is clear
that in the inner annulus, emission is brightest in the 90\degree\ sectors
centred on north and south, and faintest in the west-southwest. In the
outer annulus most of the sectors have comparable surface brightness
(though the western sectors are marginally fainter) but the south-southeast
sector is a factor $\sim$2 brighter than its neighbours (at 3$\sigma$
significance), confirming our identification of excess emission in this
direction in the image. Since the structure appears to be diffuse, it seems
likely that it is a more extended component of the galactic wind.

The 1.4~GHz radio continuum map (Figure~\ref{fig:Xopt}) shows unresolved
sources at the positions of NGC~838 and NGC~839, with low-surface brightness
diffuse emission to their south and east, extending $\sim$75\arcs\
($\sim$20~kpc) from NGC~838 and $\sim$110\arcs\ ($\sim$30~kpc) south of
NGC~839. There is no obvious source for this emission except relativistic
plasma ejected by the galactic superwinds of these two starburst galaxies,
or potentially by their AGN. Combined with the the X-ray and 610~MHz
imaging indicating that the winds bend to the east on scales of a few kpc,
this diffuse emission suggests that wind material transported out of the
galaxies is either driven eastward by the motion of the surrounding \Hi\ or
IGM, or is left behind as the galaxies move westward. We will return to the
question of interaction between the winds and the surrounding environment
in Section~\ref{sec:IGM}.

\subsection{Group-scale diffuse X-ray emission}
To investigate the properties of the diffuse intergalactic emission, we
first exclude regions around the galaxies to avoid contamination. For
NGC~838 and NGC~839 we used $\sim$25\arcs\ radius circular regions, based
on a curve-of-growth analysis designed to ensure that 95\% of
the 0.5-2~keV flux from the galaxies is excluded. Since NGC~833 and NGC~835
are interacting, with much of their gas content between the two galaxies,
and some emission from a tidal arm extending east from NGC~835, we used a
polygonal region approximating the extent of the stellar component of the
galaxies. These regions are described in more detail in Paper I. Having
excluded the galaxies, we divided the remainder of the ACIS-S3 field of
view into several large regions, based on the adaptively smoothed contours.
The upper right panel of figure~\ref{fig:Xopt} shows the rectangular North
and South Ridge regions. We defined the remainder of the S3 field
of view as the Outer region. 

The emission in all these regions is spectrally soft and extremely faint,
and probably fills the field of view. Since blank-sky fields scaled to
match the particle flux of our data may over- or under-subtract the soft
galactic foreground emission, we chose to model the background for these
regions following a method similar to that used by \citet{Snowdenetal04}
for \xmms\ data. We fit the regions simultaneously to provide the maximum
constraint on the background model.  Since the observations prior to 2013
are relatively short and have different fields of view, we only use spectra
from ObsIDs 15181, 15666 and 15667, so as to simplify the model and avoid
any uncertainties associated with changes in effective sensitivity over the
life of the ACIS instrument.

\begin{deluxetable*}{lcccccc}
\tablewidth{0pt}
\tablecaption{\label{tab:diff}Best-fitting model parameters for the diffuse low-surface brightness emission}
\tablehead{
\colhead{Region} & \colhead{Background} & \colhead{kT} & \colhead{Abund.} & \colhead{$L_{0.5-7}$} & \colhead{Surface Brightness} & \colhead{red. $\chi^2$/d.o.f.} \\ 
\colhead{} & \colhead{} & \colhead{(keV)} & \colhead{(\Zsol)} & \colhead{(10$^{38}$\ergps)} & \colhead{(10$^{38}$\ergps~arcmin$^{-2}$)} & \colhead{}\\
}
\startdata
North Ridge & Model & 0.27$\pm$0.03 & 0.05$^{+0.07}_{-0.03}$ & 94.72$^{+73.72}_{-48.13}$ & 8.63$^{+6.72}_{-4.39}$ & 1.31/2584$^\dagger$\\[+0.5mm]
            & Outer & 0.24$\pm$0.03 & 0.3$^*$ & 58.06$\pm$1.15 & 5.27$\pm$0.12 & 1.07/253 \\[+0.5mm]
            & Outer & 0.24$^{+0.03}_{-0.02}$ & $>$0.15 & 53.86$^{+1.53}_{-0.76}$ & 2.72$^{+0.15}_{-0.08}$ & 1.08/252 \\[+0.5mm]
South Ridge & Model & 0.34$^{+0.06}_{-0.03}$ & 0.05$^{+0.07}_{-0.03}$ & 190.21$^{+115.35}_{-95.49}$ & 13.18$^{+8.02}_{-6.61}$ & 1.31/2584$^\dagger$\\[+0.5mm]
            & Outer & 0.45$^{+0.13}_{-0.09}$ & 0.3$^*$ & 77.54$\pm$1.15 & 5.35$\pm$0.12 & 1.11/345 \\[+0.5mm]
            & Outer & 0.50$^{+0.15}_{-0.13}$ & 0.04$^{+0.05}_{-0.03}$ & 92.81$^\pm$1.15 & 6.45$\pm$0.12 & 1.10/344 \\[+0.5mm]
Outer & Model & 0.30$^{+0.07}_{-0.05}$ & 0.05$^{+0.07}_{-0.03}$ & 301.36$^{+290.66}_{-176.46}$ & 6.38$^{+6.15}_{-3.74}$ & 1.31/2584$^\dagger$

\enddata
\tablecomments{$^*$ Parameter fixed during fitting.\\
$^\dagger$ Fit statistic for simultaneous fit to all regions including background model.}
\end{deluxetable*}

Our spectral model consists of several components: 1) A broken power law
representing high-energy particles, which is convolved with the instrument
Response Matrix File (RMF) but not the Auxiliary Response File (ARF) since
the particles are to first order unaffected by the X-ray mirrors; 2)
Gaussians representing fluorescent emission lines within the detector, the Si K$\alpha$ and Au M$\alpha\beta$ in the case of the S3 CCD; 3) A powerlaw with $\Gamma$=1.46 and
initial normalization 8.88$\times$10$^{-7}$ per square arcminute,
representing the cosmic hard X--ray background; 4) Three thermal models
representing emission from the local hot bubble and the Galactic halo, two with
temperature 0.1~keV (one with Galactic absorption and one without) and the
third with an initial temperature of 0.25~keV; 5) A source component
consisting of an APEC thermal plasma model at the systemic redshift
($z$=0.0132) with fixed Galactic absorption. The normalisations of
components 1, 2 and 5 are allowed to fit independently for each region, but
normalisations for components 3 and 4 are linked across the regions,
scaling for area.

In addition to the \chandra\ spectra we also fit components 3 and 4 (the
X-ray foreground and background emission) to a \rosat\ All-Sky Survey
spectrum extracted from an annulus between 0.5-0.75\degree\ from the group
centroid (490-740~kpc at our adopted distance). This provides additional
constraints on the soft emission, particularly useful given the low
temperature of the group emission. Fits are carried out in the 0.5-10~keV
band, the inclusion of 7-10~keV emission helping to constrain the particle
background component in the \chandra\ data.

Table~\ref{tab:diff} shows our results. Our fits suggest the presence of weak thermal emission throughout the S3 field of view. For the set of three regions (North and South Ridge plus Outer) we experimented with thawing the normalization of the cosmic hard background component and the temperature of the 0.25~keV Galactic soft foreground. The former falls 22\% below its initial value, which we consider acceptable as we have excluded a number of bright background point sources. The latter is poorly constrained when fitted, and we therefore fix it at 0.25~keV in our final fit, which has reduced $\chi^2$=1.31 for 2584 degrees of freedom. 

As a test of the reliability of the background modelling approach, we
also fitted the spectra of the north and south ridge regions using the
spectrum of the outer region as the background. We fit each region with a
simple absorbed APEC model, and tried fits with abundance free to vary or
fixed at 0.3\Zsol. The results are listed in Table~\ref{tab:diff}. In the
north ridge the temperatures agree within the 1$\sigma$ errors with our
background modelling fit. The agreement in the south ridge is poorer, but
still at the 2$\sigma$ level. As expected, the fluxes are systematically
lower than those found from the background modelling approach, suggesting
that the background is over-subtracted owing to the presence of source flux
in the outer region. For the north ridge the abundance is poorly constrained
when fitted, owing to the small number of net counts in the spectrum after
background subtraction. In the south ridge abundance is constrained to
0.01-0.09\Zsol, but fixing it at 0.3\Zsol\ only makes the fit slightly
poorer. In general, these fits confirm the accuracy of our background
modelling approach.

To search for evidence of any more extended emission, we extracted spectra
from the S2 CCD for all three 2013 observations. The S4 CCD was not active
during the 2013 observations. S2 covers the area immediately
south-southeast of the group, but it is a front illuminated CCD and
therefore less sensitive to spectrally soft emission than S3. After
examining a heavily smoothed image, we elected to use a
$\sim$5.8$\times$8\arcm\ region excluding strips of the CCD close to the
edge of the ACIS-S array. These regions include most of the point sources
most strongly blurred by the point spread function (PSF), and could be
affected by any imperfections in the calibration of absorption by the
contaminant which has built up on the ACIS optical filters.

We excluded point sources and used the same fitting approach described
above, with additional Gaussian components to fit the Ni K$\alpha$ and Au
L$\alpha$ fluorescent emission lines which are visible on this CCD.  We
found that the background model did a good job of describing the spectrum,
with no residual features indicative of source emission. When we added a
thermal model (with abundance fixed at 0.3\Zsol) to represent diffuse IGM
emission, we found that with temperature free to vary, the best fit had
kT$<$0.11 (1$\sigma$ limit) and an extremely poorly constrained
normalization. Fixing kT at 0.3~keV we found that the normalization was
consistent with zero, with a 1$\sigma$ upper limit on surface brightness of
1.36$\times$10$^{38}$\ergps~arcmin$^{-2}$. These are very weak constraints,
and would not rule out emission consistent with the extended IGM detected
in the ``Outer'' region on S3. They do however suggest that the bright
ridge seen around the galaxies does not extend any significant distance
into the region covered by S2.

\section{Discussion}
Smoothed images of the group show extended low surface-brightness emission linking the four main galaxies and probably extending southeast beyond NGC~839. Our spectral fitting suggests that this gas has a low abundance and temperature $\sim$0.3~keV, in agreement with previous \rosat\ observations \citep{Ponmanetal96,DosSantosMamon99}. This temperature is somewhat lower than that reported from \xmm\ \citep{Belsoleetal03} except in the south ridge, where the two are comparable within errors, but this is unsurprising given the uncertain calibration of that dataset. Our low abundance of Z=0.05$^{+0.07}_{-0.03}$\Zsol\ agrees well with the \rosat\ and \xmm\ measurements \citep{DosSantosMamon99,Belsoleetal03}. 

Our \chandra\ observations seem to agree better with the \rosat\ than \xmm\
in terms of the morphology of the diffuse gas.  \citet{DosSantosMamon99}
found the diffuse emission to be clumpy and filamentary, while
\citet{Belsoleetal03} were able to model the diffuse emission as a smooth,
if highly elliptical, $\beta$-model. Belsole et al.  argued that the
superior collecting area and smaller PSF of \xmms\ allowed them to better
resolve and excise point sources, and that part of the clumpiness of the
\rosat\ image arose from point source contamination. The specific example
they raised, the C4 region identified by Dos Santos \& Mamon southwest of
NGC~833, is outside the \chandra\ field of view, so a direct comparison is
not possible. However, we find that the distribution of diffuse emission in
the S3 field of view is not consistent with a $\beta$-model. Even if
extreme ellipticities are allowed, the model overestimates the flux in the
region between NGC~835 and NGC~838, while underestimating the flux at the
ends of the ridge, particularly in the southeast corner of the field. The
curve of the ridge and the lack of a clear decline in surface brightness
southeast of NGC~839 both suggest that the gas distribution is not a simple
ellipsoid.  The disagreement with the Belsole et al. model is perhaps
unsurprising, since imaging analysis of the relatively shallow first light
\xmms\ data presented a number of difficulties, including high electronic
noise, the uncertain calibration of the EPIC-MOS cameras and the need to
exclude the then-uncalibrated EPIC-pn, scattered light from bright sources
in the field, and the difficulty of modelling the various components of the
background. We are fortunate in having deep observations from a mature,
well-calibrated instrument with superb spatial resolution.

\subsection{Physical properties of the hot intra-group medium}
\label{sec:IGM}
We can estimate the properties of the gas in the northern and southern
parts of the ridge from the results of our spectral fits, assuming a
cylindrical geometry. We approximate the north ridge as a cylinder of
length 4.3\arcm\ (70.4~kpc) and radius 1.6\arcm\ (25.9~kpc), and for the
south ridge use a length of 6.35\arcm\ (104~kpc) and radius 1.2\arcm\
(20.1~kpc). Based on the normalization of our background modelling fits to
the diffuse emission, we find that the electron number density of the
diffuse gas is 1.16$^{+1.03}_{-0.83}$$\times$10$^{-3}$\pcmcu\ and
1.16$^{+0.91}_{-0.83}$$\times$10$^{-3}$\pcmcu\ for the north and south
ridge regions respectively, neglecting any uncertainties in volume. The
bolometric luminosities for the two regions are $L_{X,{\rm
    bolo}}$=1.68$^{+1.31}_{-0.86}\times$10$^{40}$\ergps\ (north ridge) and
$L_{X,{\rm bolo}}$=3.08$^{+1.87}_{-1.55}\times$10$^{40}$\ergps\ (south).
The gas pressure in these regions is
$\sim$5-6.5$\times$10$^{-13}$~dyne~cm$^{-2}$ (equivalent to \ergpcmcu), entropy is $\sim$25-30 keV\pcmsq\
and the isobaric cooling times are $\sim$7-10~Gyr. While the low gas
entropy is comparable to that found in the cool cores of relaxed, X-ray
bright groups \citep{Sunetal09}, the cooling times are long, suggesting
that radiative cooling will probably play only a minor role in the future
development of the gas.

\begin{figure}
\includegraphics[width=\columnwidth,bb=20 205 565 745]{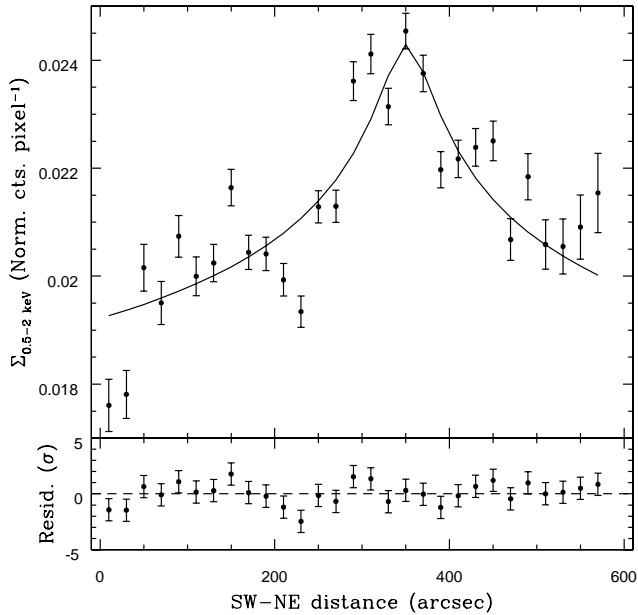}
\caption{\label{fig:SB}One dimensional exposure corrected 0.5-2~keV surface brightness profile running from southwest to northeast across the ridge of diffuse emission in HCG~16. The solid line shows the best-fitting $\beta$-model plus constant background.}
\end{figure}

The mass of gas in the two regions, excluding the denser material in and
around the galaxies, is 3.5$^{+3.1}_{-2.5}$$\times$10$^9$\Msol\ for the
north ridge and 3.1$^{+2.4}_{-2.2}$$\times$10$^9$\Msol\ for the south
ridge. This is a significant mass of gas, comparable to the overall \Hi\
deficiency estimated for the group, $\lesssim$10$^{10}$\Msol\
\citep{VerdesMontenegroetal01}. However, the spectral fits show that IGM
emission extends outside these regions, and probably outside the field of
view. Estimating the total gas mass in the system requires a model of
surface brightness. Since a 2-dimensional $\beta$-model is a poor fit to
the diffuse emission, we modelled the 1-dimensional surface-brightness
profile across the ridge. We measured the 0.5-2~keV surface brightness
using 20$\times$600\arcs\ boxes aligned at an angle of 30\degree\ (measured
anti-clockwise from west) with their long axes along the ridge. Again, the
regions used to extract spectra for the galaxies were excluded, as were
point sources outside the galaxies.The resulting profile extend from the
southwest (lower right) corner of the S3 CCD to the NE (upper right)
corner, and is shown in Figure~\ref{fig:SB}.

We modelled the exposure corrected profile with a constant background component plus a $\beta$-model. The best-fitting model is extremely flat, with core radius $R_c$=12.03\arcs$^{+57.44}_{-13.04}$ (3.28$^{+15.68}_{-3.56}$~kpc) and $\beta$=0.181$^{+0.26}_{-0.01}$. This is probably unphysical; the 1$\sigma$ upper limit on $\beta$ is more comparable to other cool groups \citep[e.g.,][]{Mulchaeyetal03}. We therefore emphasize that, while we have used this model to estimate the gas mass and luminosity of the group at large radius, the results cannot be considered as reliable measurements.

To determine how far to extrapolate this model, we must make some
assumptions about the structure of the gas, based on its likely origin. If
the gas is a virialized IGM, bound within the gravitational well of the
group, we can estimate a typical scale size for a system of this
temperature from measured scaling relations. We use the mass-temperature
relation of \citet{Sunetal09}, which was derived from a collection of
groups and clusters observed with \chandra.

For a group with kT=0.3~keV we estimate that
M$_{500}$=4.0$\times$10$^{12}$\Msol and $R_{500}$=240~kpc
($\sim$14.6\arcm). Taking a mean radius of 80\arcs\ for the two cylinders
used to model the north and south ridges, we find that extrapolating to
14.6\arcm\ would increase the gas mass by a factor $\sim$33.9 for
$\beta$=0.181 or a factor 6.2 if $\beta$ is at the 1$\sigma$ upper limit
value of 0.45. These values assume a cylindrical geometry with length
$\sim$10\arcm, so an additional factor $\sim$2 should be added if the IGM
is ellipsoidal. This gives a hot gas mass of $M_{\rm
  gas}$(R$<$$R_{500}$)$\sim$0.8-4.5$\times$10$^{11}$\Msol.

If the hot gas is associated with, and has an extent similar to, the \Hi\
filament that links the group galaxies, it is likely to be considerably
smaller, with a minor axis radius of only 4-5\arcm\ and a major axis
$\sim$20\arcm. In this case we expect an extrapolation factor of
$\sim$8-14, for a total hot gas mass of 5.2-9.4$\times$10$^{10}$\Msol,
2-3.5 times the total \Hi\ mass in the system.

A similar process of extrapolation is required to estimate the total
luminosity of the IGM. Assuming emission to be proportional to density
squared (i.e., neglecting any temperature or abundance variation) we find
that the uncertainties in the surface brightness model lead to very large
uncertainties in the total luminosity for the case of an ellipsoidal
virialized IGM, $L_{X,{\rm
    bolo}}$(R<$R_{500}$)$\simeq$1.1-8.6$\times$10$^{41}$\ergps.  For a
filamentary distribution similar to that of the \Hi, $L_{X,{\rm
    bolo}}\simeq$2.5-6.1$\times$10$^{41}$. The lower end of these estimates
are comparable to the measured luminosity in the S3 field of view without
any extrapolation, $L_{X,{\rm
    bolo}}$=1.87$^{+1.03}_{-0.66}$$\times$10$^{41}$\ergps.

\begin{figure}
\includegraphics[width=\columnwidth,bb=40 300 560 740]{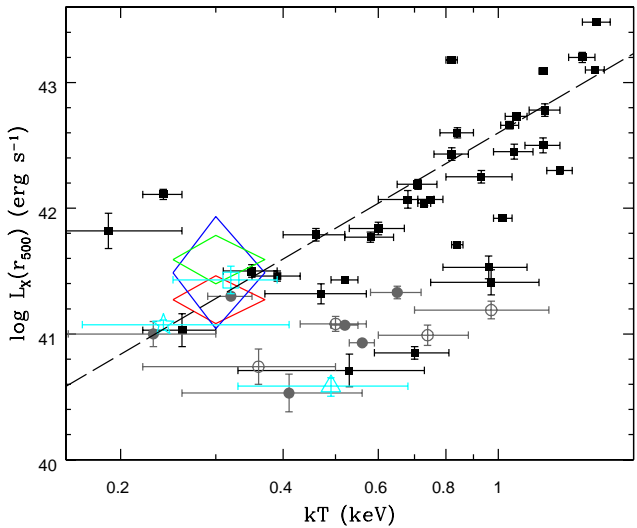}
\caption{\label{fig:LT}Luminosity and temperature for the \citet{OsmondPonman04} sample of galaxy groups, with estimates for HCG~16 overlaid. Black squares indicate the Osmond \& Ponman G sample of bona-fide groups, grey circles their H sample of groups with only galaxy-scale X-ray emission. Open circles indicate H sample systems containing only spirals. The diamonds show our own estimates using no extrapolation (red), extrapolation assuming a filamentary distribution (green) or a virialized halo (blue). Open, cyan points represent estimates by \citet[triangle]{Belsoleetal03}, \citet[star]{DosSantosMamon99} and Osmond \& Ponman (square). The dashed line shows the measured luminosity-temperature relation for groups and clusters from \citet{Eckmilleretal11}, a study based on \chandra\ data for systems with kT$>$0.5~keV.}
\end{figure}

Figure~\ref{fig:LT} shows a comparison of our luminosity estimates with the luminosity-temperature distribution of a sample of X-ray bright groups from \citet{OsmondPonman04}. Their sample was divided into confirmed groups with diffuse X-ray emission extending $>$65~kpc (their G subsample, black squares) and less luminous systems which only host galaxy-scale extended emission (H sample, grey circle). Apart from HCG~16, none of their G groups were spiral-only systems. The H sample did include four spiral-only systems, which fall on the lower edge of the luminosity-temperature (L-T) distribution. All three of our estimates fall within the scatter of the distribution, which is large at low temperatures, though the upper bound of our luminosity estimate assuming a fully virialized halo is at the upper edge.

For comparison, we also show previous estimates of the luminosity and temperature of HCG~16, corrected to our adopted distance. However, we note that these values are not truly comparable, owing to the different apertures used and different approaches to exclusion of galaxy emission and point sources. The \xmms\ estimate of \citet{Belsoleetal03} differs from ours by the largest factor, with a marginally higher temperature and significantly lower luminosity. The \rosat\ estimate of \citet{DosSantosMamon99}, using their fit with abundance fixed at 0.1\Zsol, is in agreement with our unextrapolated luminosity estimate, while the Osmond \& Ponman estimate is comparable to all three of our luminosity estimates. In summary, our \chandra\ measurements are consistent with the measured luminosity-temperature relation for groups, within the large observed scatter.

\subsection{The origin of the hot IGM}
Three possible sources of the hot gas in the intra-group medium of HCG~16
can be suggested: 1) Primordial gas which has fallen into the group
potential and been shock heated to the virial temperature; 2) Shock-heated
\Hi\ from the large-scale filament, heated by a high-speed collision as in
Stephan's Quintet; 3) Material ejected from the member galaxies by galactic
winds during phases of intense star formation. Perhaps the most notable
feature of the IGM is the partial correlation between the positions of the
hot and cold gas in the group, with the X-ray and \Hi\ emission co-located
around the starburst galaxies NGC~838 and NGC~839. This, combined with the
ridge morphology of the hot gas, strongly suggests a connection between the
two different gas phases, at least in the group core.

In most relaxed, X-ray bright groups, the IGM produces an elliptical or
circular surface brightness distribution \citep[see
e.g.,][]{Mulchaeyetal03}. Ridge-like structures linking galaxies and
superimposed on an underlying ellipsoidal structure are not unknown
\citep[e.g., the NGC~5171 group,][]{Osmondetal04} but are uncommon. The
only other spiral-only group detected in the X-ray band, SCG0018-4854, is
too X-ray faint to allow a detailed morphological study
\citep{Trinchierietal08}, but Stephan's Quintet, whose galaxy population is
only slightly more evolved than HCG~16, has an apparently relaxed
large-scale IGM outside the complex group core
\citep{Trinchierietal05,OSullivanetal09}. The S3 field of view is too small
to allow us to determine whether the large-scale X-ray emission outside the
ridge is relaxed or not, but the failure to fit a $\beta$-model to the
ridge rules out a simple ellipsoidal model such as that used by
\citet{Belsoleetal03}.  Certainly the ridge is the dominant component of
the IGM in the \chandra\ field of view, and the distribution of emission
outside the ridge is very flat.

We can consider the stability and likely lifespan of the ridge structure as
a constraint on its likely origin. Taking a radius of 25~kpc and a typical
temperature of 0.3~keV, the sound crossing time of the ridge is $\sim$112~Myr.
The ridge has a higher density (and therefore pressure) than the
surrounding IGM, and should therefore expand and disperse over the course
of a few hundred Myr, unless it is somehow confined. An extended
filamentary structure like the ridge could not be formed by the relaxed
gravitational potential of a virialized group, but might be temporarily
formed by the close association of the group member galaxies, with their
individual dark matter halos helping to retain some of the gas around the
galaxies. The lifetime of the southern ridge might be extended by the ongoing
starburst winds from NGC~838 and NGC~839, which could inject new
higher-density gas as older material expands outward. The southern ridge
has a surface brightness a factor 1.5 greater than the northern ridge, and
starburst winds may explain this difference. The north ridge has only marginally higher surface brightness than its surroundings (8.63$^{+6.71}_{-4.39}$$\times$10$^{38}$\ergps~arcmin$^{-2}$ compared to 6.38$^{+6.15}_{-3.74}$$\times$10$^{38}$\ergps~arcmin$^{-2}$ for the IGM) and this may be related to the lack of recent star formation in NGC~833 and NGC~835. 

Another possibility is that our spectral fits to the diffuse X-ray emission
are misleading, and that the ridge is actually a region of enhanced
abundance rather than enhanced density. Galaxy winds could have enriched
the IGM around the galaxies, producing higher surface brightness through
enhanced line emission. Quite small increases in abundance can
significantly increase surface brightness at the low temperatures observed
in the group, and doubling the observed 0.05\Zsol\ abundance could explain
the factor $\sim$1.3 change in mean surface brightness across the ridge.
This scenario is again consistent with a higher surface brightness in the
southern ridge, since its enrichment is ongoing. However, it is difficult
to see how the ridge can be a product of enrichment alone; the same galaxy
winds which transport heavy elements out into the ridge also bring higher
density gas. It seems likely that all of these possibilities contribute,
and that the ridge is a temporary structure formed by the gravitational
interaction of the major group members, with starburst winds helping to
boost its surface brightness, extent and lifetime.

\subsubsection{A virialized halo}

The estimated gas mass in the IGM for a fully virialized 0.3~keV system
implies a (hot) gas mass fraction within $R_{500}$ of $f_{\rm gas}^{\rm
  hot}$=0.02-0.11, a stellar fraction of $f_{*}$=0.08, and a baryon
fraction of $f_{\rm baryon}>$0.11-0.20, taking
$M_{500}$=4$\times$10$^{12}$\Msol\ from the M:T relation of
\citet{Arnaudetal05} and neglecting the contribution of dwarf galaxies and
any intergalactic stellar component. The lower bound of this range is just
comparable with the upper limit of scatter seen in the lowest mass groups
and poor clusters for which accurate measurements have been made
\citep[e.g.,][]{Sandersonetal13,Gonzalezetal13}, while the upper bound
exceeds the universal baryon fraction. We also note that the Sanderson et
al. and Gonzalez et al. baryon fraction measurements include a significant
contribution from intra-cluster stars, which our estimate neglects.
Including this component in HCG~16 would increase the baryon fraction,
making the agreement with other groups even poorer. Our alternative
estimate, assuming the hot gas distribution is comparable to that of the
\Hi, would give a somewhat more reasonable $f_{\rm baryon}>$0.12-0.13, again neglecting intra-cluster stars.

Similarly, the stellar fraction is $\sim$50\% of the universal baryon
fraction, considerably higher than observed in more massive groups and
clusters \citep{Sandersonetal13,Gonzalezetal13}. Stellar fraction is
expected to peak in systems with mass M$_{500}\sim$10$^{12}$\Msol\
\citep{Leauthaudetal12}, but $f_{*}\sim$0.5 would place HCG~16 at the
extreme upper edge of the likely range.

It therefore seems unlikely that the IGM of HCG~16 (or at least that part
of it we can observe) is relaxed, or that it has formed entirely through
gravitational infall. The \chandra\ results thus support the conclusions
drawn from the \rosat\ data by \citet{Ponmanetal96} and
\citet{DosSantosMamon99}. This raises the question of why the group falls
on or above the L-T relation. We might expect that the IGM in a virialised
group would be hotter than in one that has yet to virialize,
since the fully collapsed system is denser and better able to compress the
gas. However, if a large fraction of the gas is heated by non-gravitational
processes such as star formation or shocks, its temperature will depend on
the balance between those heating processes and losses from radiative
cooling and gas mixing.

\subsubsection{Shock heating}

Shock heating of the \Hi, as observed in Stephan's Quintet, would require a
high velocity collision. The line-of-sight velocities of the five major
galaxies in HCG~16 cover a range of only 227\kmps, with the largest
difference, between NGC~833 and NGC~835, probably arising from their
interaction. However, if we consider that the age of the starbursts in
NGC~838 and NGC~839 are likely to correspond to tidal encounters with
NGC~848, we can estimate the velocity of NGC~848 in the plane of the sky to
be 455-475\kmps. A collision at this velocity would produce a strong shock
in a 100K \Hi\ cloud; a head-on collision would raise the temperature of
the \Hi\ to $\sim$0.42~keV ($\sim$5$\times$10$^6$K). This is quite similar
to the observed temperature of the IGM, suggesting that if shock heating is
the primary source of gas, cooling must have had little impact since the
shock occurred. 

NGC~848 would also have caused a weak shock in any hot gas it encountered
while passing through the group core. A temperature of 0.3~keV implies a
sound speed of $\sim$220\kmps\ in the IGM, so NGC~848 would have produced a
Mach$\sim$2.1 shock, a temperature increase of factor $\sim$2.2 and a
density increase of factor $\sim$2.4. A weak shock would therefore produce
a large increase, of a factor $\sim$10, in the bolometric X-ray luminosity
of any hot gas in the system. The current data are not capable of
differentiating between a ridge primarily formed from shocked \Hi, and one
with a significant contribution from shocked hot gas, but the effects of
such shocks in compact groups would make an interesting subject for
investigation with future numerical simulations.

Only the S1 CCD of ObsID 923 covers NGC~848, where any shock front would
currently be located, and the short exposure of the observation and large
off-axis angle mean that it lacks the depth and spatial resolution to
detect such a feature. None of the other indicators of a shock observed in
Stephan's Quintet (optical and infra-red line emission, radio continuum
emission, star formation outside the major galaxies) are present in HCG~16,
though most of them are short lived and would likely have faded over the
$\sim$500~Myr since NGC~848 passed the other galaxies.  The large mass of
\Hi\ observed in NGC~848 and in the large-scale filament strongly suggests
that shock heating can only have affected a small fraction of the gas. We
can place a limit on the mass of \Hi\ that could have been shock heated of
$\lesssim$1.1$\times$10$^{10}$\Msol, based on the \Hi\ deficiency of the
group \citep{VerdesMontenegroetal01}. At best, shock heated \Hi\ may
therefore account for $\sim$20\% of the IGM.

\subsubsection{Galaxy winds}

The third possible source of gas is the star formation driven galactic
winds in the member galaxies. Our observations of NGC~838 and NGC~839 show
that the winds of these starburst galaxies have temperatures of
0.8-0.9~keV, with some indication of a temperature decline in their outer
parts (see Paper I). Although the radiative cooling times of the hot gas in
the winds are $>$1~Gyr, we might expect more rapid cooling to be caused by
mixing with cold gas entrained in the winds or encountered as wind material
interacts with the \Hi\ filament. This might explain the reduction in
temperature by a factor of 3 from the galaxy winds to the IGM. It might
also explain the marginally higher temperature of the diffuse gas in the
south ridge (0.34~keV) where the two starburst galaxies reside, compared to the north ridge, which is occupied by AGN-dominated systems (0.27~keV).

Our estimate of the mass of gas which may have been ejected by NGC~838 and
NGC~839 (9.5$\times$10$^9$\Msol) exceeds the mass of hot gas in the
southern ridge by a factor $\sim$3, suggesting that it could be largely
formed from ejected wind material. For the northern ridge to have formed in
a similar way, the star formation rates of NGC~833 and NGC~835 would have
to have been significantly greater in the past. Given the limits on recent
star formation from our stellar population modelling of NGC~833 (see paper
I) their starburst periods would need to be significantly older, perhaps
triggered by an initial tidal encounter between the two galaxies. From the
velocity estimated above, NGC~848 seems likely to have passed NGC~833/835
$\sim$600~Myr ago, and a starburst triggered then would be clearly detected
in the SDSS spectrum of NGC~833. If we assume that NGC~833 and NGC~835 did
go through a superwind phase, then a total of
$\sim$2$\times$10$^{10}$\Msol\ of hot gas would have been produced by the
four original group members, $\sim$20-40\% of our estimated IGM mass. For a
Salpeter initial mass function, stellar populations are expected to lose
$\sim$30\% of their mass over a Hubble time \citep{White91,Davidetal91}.
For the four original members this puts a strong upper limit of
$\sim$9$\times$10$^{10}$\Msol\ of hot gas which could be produced from the
galaxies, roughly five times the amount we expect to have been ejected by
stellar winds.

If wind material does make up a significant fraction of the IGM, we might
expect to see higher abundances. The abundance measured in NGC~838 is only
$\sim$0.16\Zsol, but is probably biased low because even the hot phase of
the wind is multi-temperature and because of mass loading; as the enriched
hot gas produced by stellar winds and supernovae flows out of the galaxy it
entrains less enriched neutral hydrogen which effectively dilutes the
metallicity. This is likely to be the case in the IGM as well since we
observe \Hi, X-ray and radio continuum emission from the same regions,
indicating that the IGM is multi-phase.

There is also the question of whether hot wind material can escape the
immediate neighbourhood of the starburst galaxies and diffuse out through
the surrounding \Hi. The diffuse radio continuum emission demonstrates that
gas from the winds can reach distances of 20-30~kpc, and the X-ray and
radio morphology of the winds suggests that while they may be affected by
the motion of the \Hi\ relative to the galaxies, they are not completely
confined by it. It seems likely that the large-scale \Hi\ filament is
actually a complex of smaller structures, interspersed with hotter
material. While some mixing must take place, resulting in heating of the
\Hi\ and cooling of the hot X-ray emitting gas, the presence of the X-ray
ridge demonstrates that this is not an efficient process. Hot gas clearly
coexists with the \Hi\ over significant timescales.
\citep{Borthakuretal10} have estimated that \Hi\ clouds of radius
$\geq$200~pc (0.73\arcs) can survive for a few hundred Myr in a hot IGM, so
this is not implausible.

We can question whether the energy available for star formation is
sufficient to heat this mass of gas to its current temperature. The total
energy of the gas in the north and south ridges is
$\sim$4$\times$10$^{57}$~erg. The number of core collapse supernovae
expected per unit star formation is 0.01-0.015 \citep{Botticellaetal12},
and we assume the standard value for the energy available from each
supernova, 10$^{51}$~erg. This suggests that over the 500~Myr (400~Myr)
timescale of star formation in NGC~838 (NGC~839), a star formation rate of
0.25-0.35\Msolpyr\ (0.35-0.5\Msolpyr) would be needed to heat the gas
through supernovae, assuming perfect efficiency and no losses. Accounting
for radiative losses from the hot gas, the much higher star formation rates
for the two galaxies measured in paper I suggest that efficiencies of 2-4\%
would be sufficient to heat the gas in the ridge to its observed
temperature. 

However, the enrichment of the IGM by heavy elements in the galactic winds
must also be considered. If the abundance of the ridge is truly 0.05\Zsol,
this implies that the ridge contains $\sim$10400\Msol\ of Fe. Production of
heavy elements is likely to be dominated by core collapse supernovae in the
superwind galaxies, so adopting an Fe yield of 0.07\Msol\ per supernova
\citep{Finoguenovetal00}, we would require only $\sim$1.5$\times$10$^5$
supernovae to produce the observed enrichment. This is insufficient to heat
the ridge gas to the observed temperature. Even if the abundance were
0.3\Zsol, we would still only require $\sim$9$\times$10$^5$ supernovae,
between 1/5 and 1/7 of the number required to heat the gas in the ridge. 

If we use our estimate of the number of supernovae required to heat the
gas, we find that the expected Fe abundance is $\sim$1.3\Zsol. However, we
only expect galactic winds to have contributed 20-40\% of the gas in the
ridge.  If the remaining 60-80\% is low-abundance, unenriched gas heated by
some other process (gravitational collapse and/or shocks) mixing will
reduce the expected abundance to $\sim$0.25-0.5\Zsol. The lower end of this
range could be consistent with our abundance measurements, once the effect
of biases arising from the multiphase nature of the IGM are taken in to
account. We therefore conclude that it is plausible both energetically and
in terms of mass, that galactic winds from the member galaxies have played
a significant part in forming the IGM we observe in HCG~16, though they
cannot be the only source of hot gas.

\section{Conclusions}
\label{sec:conc}

HCG~16 is one of only two spiral-only groups known to support an X-ray
luminous hot IGM. As such, it provides a unique view of the early stages of group evolution, in which hot and cold gas phases coexist and tidal interactions have begun to reshape the galaxy population. The possibility that dynamical interactions and galaxy winds may play a significant role in the build up of the hot IGM is of obvious importance to our understanding of the development of galaxy groups and their member galaxies. We have analysed new deep \chandra\ observations of HCG~16 with the goal of determining the properties and likely origin of its hot gaseous halo. Our results can be summarised as follows:

\begin{enumerate}

\item We confirm the presence of an extended hot IGM in HCG~16 with a
  temperature $\sim$0.3~keV and abundance 0.05$^{+0.07}_{-0.03}$\Zsol. Hot
  gas is found to extend through the ACIS-S3 field of view, with the
  brightest emission forming a ridge linking the four original group
  members and extending to the southeast, in rough agreement with previous
  \rosat\ and \xmm\ results. The ridge partly overlaps the \Hi\ tidal
  filament that links the five major galaxies, particularly in the region
  around NGC~838 and NGC~839, suggesting that the two gas phases are
  intermingled. The ridge contains 6.6$^{+3.9}_{-3.3}$$\times$10$^9$\Msol\
  of gas with an entropy $\sim$25-30~keV~cm$^2$. This mass is similar to
  the \Hi\ deficiency of the group, while the entropy is comparable to the
  limit below which cooling is thought to fuel star formation and nuclear
  activity in the cooling flows of galaxy clusters \citep{Voitetal08}.
  However, the cooling time of the gas is long, 7-10~Gyr, suggesting that
  radiative cooling is unlikely to be an important factor for the hot IGM.

\item We consider three possible mechanisms which may have contributed to
  the formation of the IGM in HCG~16. The correlated X-ray and \Hi\
  morphology suggests that in the group core neither the hot nor the cold
  gas component is relaxed, while the radio and X-ray morphologies of the
  galactic winds provide evidence that hot gas can escape the galaxies and
  mix into the IGM despite the surrounding cold \Hi. The assumption that the
  group is virialized with an IGM formed through gravitational accretion
  leads to unrealistically high gas and baryon fraction estimates. A more
  limited gas distribution similar to that of the \Hi\ implies a hot gas
  mass of 5.2-9.4$\times$10$^{10}$\Msol\ and luminosity $L_{X,{\rm
      bolo}}$=2.5-6.1$\times$10$^{41}$\ergps. We consider the possibility
  that part of the hot IGM may have been formed through the shock heating
  of neutral hydrogen as NGC~848 passed through the group core, but while
  we find it likely that NGC~848 was supersonic during this passage, there
  is no direct evidence of such a shock. Taking our rates of outflow for
  hot gas in the winds of NGC~838 and NGC~839 from Paper I, we find that
  they may have ejected $\sim$2$\times$10$^{10}$\Msol\ of gas since their
  starbursts began. If NGC~833 and NGC~835 underwent similar starburst
  phases at an earlier period, galactic winds could have contributed
  20-40\% of the hot gas observed in the IGM.

\end{enumerate}  

We therefore conclude that starburst winds have played a significant role in the development of the IGM in HCG~16, alongside gravitational infall and possibly collisional shock heating of \Hi. Given the very small numbers of low-mass, spiral-rich groups at this early evolutionary stage in which a hot diffuse IGM is known to exist, it is difficult to judge how widely applicable this result is to the general population. However, the importance of understanding the process by which the IGM is formed and its relationship to other aspects of group evolution and structure formation makes it clear that further studies of similar systems are required.

\acknowledgements The authors thank L. Verdes-Montenegro for providing her
VLA \Hi\ map of the group, and the anonymous referee for helpful comments on the paper. Support for this work was provided by the
National Aeronautics and Space Administration (NASA) through Chandra Award
Number G03-14143X issued by the Chandra X-ray Observatory Center (CXC),
which is operated by the Smithsonian Astrophysical Observatory (SAO) for
and on behalf of NASA under contract NAS8-03060.  SG acknowledges the
support of NASA through the Einstein Postdoctoral Fellowship PF0-110071
awarded by the CXC, and this research has made use of data obtained from
the Chandra Data Archive and software provided by the CXC in the
application packages CIAO, ChIPS, and Sherpa, as well as SAOImage DS9,
developed by SAO. We thank the staff of the GMRT for their help during
observations. GMRT is run by the National Centre for Radio Astrophysics of
the Tata Institute for Fundamental Research. We acknowledge the usage of
the HyperLeda database (http://leda.univ-lyon1.fr).

\textit{Facilities:} \facility{CXO} \facility{VLA} \facility{GMRT}

\bibliographystyle{apj}
\bibliography{../paper}

\end{document}